\documentclass[twocolumn,showpacs,prb,aps,amsmath,superscriptaddress]{revtex4}

\usepackage{graphicx}
\usepackage{dcolumn}
\usepackage{bm}
\usepackage{amssymb}

\begin{document}

\title{Influence of photon-assisted tunneling on heat flow in a normal metal--
superconductor tunnel junction}

\author{Nikolai B. Kopnin}
\email{kopnin@boojum.hut.fi} \affiliation{Low Temperature
Laboratory, Helsinki University of Technology, P.O. Box 2200,
02015 TKK, Finland} \affiliation{L. D. Landau Institute for
Theoretical Physics, 117940 Moscow, Russia}
\author{Fabio Taddei}
\affiliation{NEST CNR-INFM and Scuola Normale Superiore, I-56126 Pisa, Italy}
\author{Jukka P. Pekola}
\affiliation{Low Temperature Laboratory, Helsinki University of
Technology, P.O. Box 2200, 02015 TKK, Finland}
\author{Francesco Giazotto}
\email{giazotto@sns.it}
\affiliation{NEST CNR-INFM and Scuola Normale Superiore, I-56126 Pisa, Italy}


\begin{abstract}
We have investigated theoretically the influence of an AC drive on heat
transport in a hybrid normal metal - superconductor tunnel junction
in the photon-assisted tunneling regime. We find that the useful
heat flux out from the normal metal is always reduced as compared to
its magnitude under the static and quasi-static drive conditions.
Our results are useful to predict the operative conditions of AC driven superconducting electron refrigerators.
\end{abstract}


\maketitle
\section{Introduction}
Photon-assisted tunneling (PAT) has been discussed in literature for
almost half a century by now
\cite{dayem,cook,lax,hamilton,kommers,prober,vaknin,yu,wyder,habbal,mooij,leone,uzawa,tien,tucker,tucker2,sweet,zimmermann}.
This phenomenon arises when a relatively high frequency field is
applied across a tunnel junction whose DC current-voltage
characteristics are highly non-linear. The radiation field is,
however, slow enough to guarantee adiabatic evolution of the energy
levels of the electrons. A typical system to observe PAT is a SIS
tunnel junction, with superconducting (S) leads and a tunnel barrier
(I) in between. Even though the system as such is a Josephson
junction for Cooper pairs, PAT deals with the influence of the
radiation on quasiparticle tunneling. Our system of interest here is
a NIS tunnel junction, where one of the conductors is a normal metal
(N). Such junctions exhibit highly nonlinear current-voltage
characteristics at low temperatures, and normally the current is due
to quasiparticles only. NIS-junctions are known to have peculiar
heat transport properties under the application of a DC bias
voltage \cite{giazotto06,bardas,leivo,nahum,clark,clark2,savin,leoni,fs,magbar,fis},
or ("quasi-static") AC radiation of relatively low
frequency in form of either periodic or stochastic drive
\cite{pekola07a,saira,pekola07b}. Specifically, it is
possible to find operation regimes where the normal metal is
refrigerated and the superconductor is overheated, and in some
special situations the opposite can occur as well. The question
remains whether and under what conditions the relatively high
frequency radiation responsible for PAT would either enhance or
suppress the thermal transport in the NIS-system. In this paper we
show that the influence of PAT, as compared to static and
quasi-static AC drive conditions, is to decrease the refrigeration
of the normal conductor, and also to change, usually to increase,
the magnitude of heat dissipation in the superconductor. Although
these results are somewhat unfortunate for high-frequency
applications of NIS-junctions, they are, however, useful in finding
operating conditions, for instance, for AC driven electronic
refrigerators \cite{pekola07a,saira}.

The paper is organized as follows. In Sec.~\ref{sec:model} we
describe the
theoretical framework
together with the discussion of the conditions of its validity. In
particular, in Sec.~\ref{sec:theory-results} we present our
analytical results for the heat and charge currents. In
Sec.~\ref{sec:result} we show and discuss the results. Finally,
our conclusion are drawn in Sec.~\ref{sec:conc}.

\section{Model and formalism}
\label{sec:model}

\begin{figure}[t!]
\includegraphics[width=0.8\columnwidth,clip]{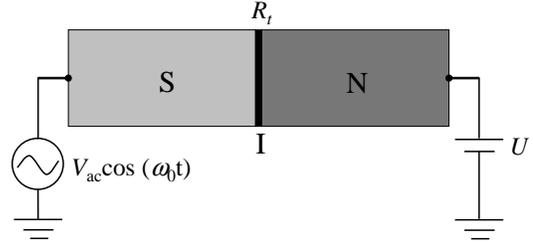}
\caption{The system under investigation is composed of a
superconductor (S) tunnel-coupled to a normal metal (N)
layer through an insulating barrier (I) of resistance $R_t$. The
superconductor is AC voltage biased with
$V_{\text{ac}}\cos(\omega_0t)$, while the N electrode is biased
with a static voltage $U$. Both electrodes are assumed to be in
thermal equilibrium.} \label{fig1}
\end{figure}

The system under investigation consists of superconducting (S) and
a normal (N) electrode tunnel-coupled through an insulating
barrier (I) of large resistance $R_t$. An AC voltage bias
$\varphi_S$, of frequency $\nu_0=\omega_0/2\pi$ and amplitude
$V_{\text{ac}}$, is applied to the S electrode, while a static
voltage $\varphi_N=U$ is applied to the N contact. The total
voltage across the junction is $\varphi_N-\varphi_S=U-V_{ac}\cos
\omega_0 t$. One could, of course, consider both the AC and DC
voltage to be applied to the normal lead, instead. However, we
choose the setup as shown in Fig.~\ref{fig1} to directly
demonstrate equivalence of the two connections when one of the
leads is in the superconducting state.

In the tunnelling limit with large resistance $R_t$ the currents
through the contact are small. If the AC frequency is small
compared to the superconducting gap, $\omega_0\ll \Delta$, the
deviation from equilibrium in each lead is negligible. In
particular, the equilibrium is preserved with respect to the
superconducting chemical potential $\mu_S$ in the S electrode
(which has dimensions much bigger than the branch-imbalance relaxation
length). This leads to the standard
assumption~\cite{tien,tucker,tucker2}
\begin{equation}
\frac{\hbar}{2} \frac{\partial \chi}{\partial t}\equiv \mu_S
=-e\varphi_S \ . \label{chem-potential}
\end{equation}
where $\chi$ is the order parameter phase.

In the case of equilibrium described by Eq.
(\ref{chem-potential}), the order parameter has the form
\begin{equation}
\Delta ({\bf r},t) =\Delta_0 ({\bf r}) \exp\left(
\frac{2i}{\hbar}\int_0 ^{t} \mu_s\, dt^\prime \right) \ .
\label{OP}
\end{equation}
It is convenient to start with the Bogoliubov--de Gennes equation
(BdGE) for the eigen-functions of the system,
\begin{eqnarray*}
i\hbar \frac{\partial u}{\partial t} &=&(\hat H_0 +e\varphi_S )u
+\Delta v \ , \\
i\hbar \frac{\partial v}{\partial t} &=&-(\hat H_0^* +e\varphi_S
)v +\Delta ^* u \ ,
\end{eqnarray*}
where $H_0$ is the normal-state Hamiltonian. Solutions to the BdGE
have the form
\begin{eqnarray}
u({\bf r},t) &=& u_0({\bf r}) e^{ -iEt/\hbar +(i/\hbar)\int
^{t} \mu_s\, dt^\prime } \ , \label{t-dep-u}\\
v({\bf r},t) &=& v_0({\bf r}) e^{ -iEt/\hbar -(i/\hbar)\int ^{t}
\mu_s\, dt^\prime } \ , \label{t-dep-v}
\end{eqnarray}
where $u_0({\bf r})$ and $v_0({\bf r})$ satisfy the BdGE in the absence of
the applied potential ($\varphi_S=0$)
\begin{eqnarray*}
E u_0 &=&\hat H_0 u_0
+\Delta ({\bf r})v_0 \ , \\
E v_0 &=&-\hat H_0^* v_0 +\Delta ^*({\bf r}) u_0 \ .
\end{eqnarray*}

Using the standard approach we define the retarded (R) and
advanced (A) Green function, which can be written as a matrix in
Nambu space:
\[
\hat G^{R(A)} = \left(\begin{array}{cc} G^{R(A)} & F^{R(A)}\\ -F^{R(A)\dagger} & \bar
G^{R(A)}\end{array}\right)\ ,
\]
where $F^{R(A)}$ refer to the anomalous Gorkov function. Since
these functions are statistical averages of the particle field
operators which can be decomposed into the wave functions Eqs.
(\ref{t-dep-u}) and (\ref{t-dep-v}), the retarded and advanced
Green functions with help of Eqs.~(\ref{t-dep-u}) and
(\ref{t-dep-v}) take the form
\begin{eqnarray*}
G^{R(A)}(t_1,t_2) &=& G^{R(A),0}(t_1,t_2)\, e^{i/\hbar (\int^{t_1} \mu_s \,
dt^\prime -\int^{t_2} \mu_s \, dt^\prime )}\ , \label{G1}\\
\bar G^{R(A)}(t_1,t_2) &=& \bar G^{R(A),0}(t_1,t_2) \, e^{ -i/\hbar (\int^{t_1}
\mu_s \, dt^\prime +\int^{t_2} \mu_s \, dt^\prime) } , \label{barG1}\\
F^{R(A)}(t_1,t_2) &=& F^{R(A),0}(t_1,t_2) \, e^{ i/\hbar (\int^{t_1} \mu_s \,
dt^\prime +\int^{t_2} \mu_s \, dt^\prime) } \ , \label{F1}
\end{eqnarray*}
where $G^{R(A),0}$, $F^{R(A),0}$ refer to $\varphi_S=0$.

If the AC voltage is applied to the superconductor
\begin{equation}
\mu_s =-eV_{\text{ac}} \cos \omega_0 t\ . \label{mu-s}
\end{equation}
Using the identity
\begin{equation}
e^{ -\frac{i}{\hbar}\int_0^t eV_{\text{ac}}\cos \omega_0 t^\prime
\, dt^\prime } =\sum_{n=-\infty}^{n=+\infty} J_n(\alpha
)e^{-in\omega_0 t} \label{expansion1}
\end{equation}
where $\alpha =eV_{\text{ac}}/\hbar \omega_0$, and $J_n$ is the $n$th order
Bessel function. The Green functions in the frequency
representation take the form
\[
G^{R(A)}_{\epsilon ,\epsilon-\hbar\omega}
=\sum_{n,m}J_n(\alpha)J_{m}(\alpha) G^{R(A),0}_{\epsilon
-n\hbar\omega_0, \epsilon -\hbar\omega -m\hbar\omega_0 }\ .
\]
If $\varphi_S=0$, there is no time-dependence and
$G^{(0)}_{\epsilon_1,\epsilon_2}=2\pi \hbar\delta
(\epsilon_1-\epsilon_2)G_{\epsilon_1}$. Here and in what follows
one frequency subscript refers to the static Green function.

The semi-classical Green functions are defined as the Green
functions in the momentum representation integrated over the
energy variable $\xi _p = p^2/2m -E_F$,
\[
\hat g_{\epsilon_1,\epsilon_2} = \int_{-\infty}^{+\infty} \hat
G_{\epsilon_1,\epsilon_2}({\bf p} ,{\bf p}-{\bf k})\, \frac{d\xi
_p }{\pi i} \ .
\]
Under the AC drive we thus have
\begin{eqnarray}
g^{R(A)}_{\epsilon ,\epsilon-\hbar\omega} &=&\sum_{n,k}2\pi \delta
(\omega -k\omega_0)J_n(\alpha)J_{n-k}(\alpha) g^{R(A)}_{\epsilon
-n\hbar\omega_0 } , \; \label{g-omega}\\
f_{\epsilon,\epsilon-\hbar\omega }^{R(A)} &=&\sum_{n,m} 2\pi
\delta (\omega -k\omega_0)
J_n(\alpha)J_{k-n}(\alpha)f^{R(A)}_{\epsilon-n\hbar\omega_0} .\;
\label{f-omega}
\end{eqnarray}

For the Keldysh functions we use the standard representation
\cite{LO,kopnin} in terms of $f_1$ and $f_2$ which are the
components of the distribution function respectively odd and
even in $(\epsilon, {\bf p})$. In Nambu space,
\begin{eqnarray*}
\hat g^K_{\epsilon _1 , \epsilon _2}=\int_{-\infty}^{+\infty}
\frac{d\epsilon^\prime}{2\pi \hbar} \left[\hat g^R_{\epsilon _1 ,
\epsilon ^\prime}\left(f_{1, \epsilon^\prime, \epsilon _2}+ \hat
\tau_3 f_{2,\epsilon^\prime,\epsilon _2}\right)\right.
\\
\left. -\left(f_{1, \epsilon _1, \epsilon^\prime,}+\hat \tau_3
f_{2,\epsilon _1, \epsilon^\prime,}\right)\hat g^A_{\epsilon
^\prime , \epsilon _2}\right]
\end{eqnarray*}
where
\[
\hat \tau_3 = \left(\begin{array}{cc} 1 & 0\\
0 & -1\end{array}\right)\ .
\]
In what follows we omit the integration limits if the integration
is extended over the infinite range. With Eqs.\ (\ref{g-omega}),
(\ref{f-omega}) the Keldysh Green functions take the form
\begin{eqnarray}
g^K_{\epsilon,\epsilon-\hbar\omega}\!\! &=&\!\! \sum_{n,k}2\pi
\delta (\omega
-k\omega_0)J_n(\alpha)J_{n-k}(\alpha)\nonumber \\
&&\!\! \times [g^{R}_{\epsilon -n\hbar\omega_0 }-g^{A}_{\epsilon
-n\hbar\omega_0 }]\nonumber \\
&&\times \left[f_{1}(\epsilon -n\hbar\omega _0) + f_{2}(\epsilon
-n\hbar\omega
_0)\right] , \nonumber \\
f_{\epsilon,\epsilon-\hbar\omega }^K\!\! &=&\!\! \sum_{n,m} 2\pi
\delta (\omega
-k\omega_0) J_n(\alpha)J_{k-n}(\alpha)\nonumber \\
&&\!\! \times \left[f^{R}_{\epsilon-n\hbar\omega_0}(f_{1,\epsilon
-n\hbar\omega_0}-
f_{2,\epsilon -n\hbar\omega_0})\right. \nonumber \\
&&\!\! \left. - (f_{1,\epsilon -n\hbar\omega_0}+f_{2,\epsilon
-n\hbar\omega_0})f^{A} _{\epsilon-n\hbar\omega_0}\right] ,
\label{F-K}
\end{eqnarray}
where the distributions $f_{1}$ and $f_{2}$ in the superconductor
refer to the state with $\varphi_S=0$. Equations for $\bar
g^{R(A)}$ and $\bar g^K$ are obtained from the corresponding
equations for $g^{R(A)}$ and $g^K$ by substituting $g\rightarrow
\bar g$, $\omega _0 \rightarrow -\omega_0$, and $f_2 \rightarrow -
f_2$.

These solutions describe a quasi-equilibrium state with a
time-dependent chemical potential Eq. (\ref{mu-s}). In the limit
$\omega_0 \rightarrow 0$, using Eq.~(\ref{expansion-ident}), we
have $ g^{R(A,K)}_{\epsilon}\rightarrow g^{R(A,K)}_{\epsilon
+\mu_S } $, $\bar g^{R(A,K)}_{\epsilon}\rightarrow \bar
g^{R(A,K)}_{\epsilon -\mu_S } $ which agrees with the
constant-voltage limit \cite{VHK05}.

Here we need an obvious remark. It can be shown (see Appendix
\ref{app}) that for $\mu_s$ satisfying Eq.\ (\ref{mu-s})
\begin{eqnarray}
&&\sum _{n,k}2\pi \delta (\omega
-k\omega_0)J_n(\alpha)J_{n-k}(\alpha) \Phi(\epsilon-n\hbar\omega_0
)\nonumber \\
&=&\int \Phi (\epsilon +\mu_s (t)) e^{i\omega t}\, dt
\label{expansion-ident}
\end{eqnarray}
for any function $\Phi (\epsilon)$ which has no singularities. For
a function with singularities (or large higher-order derivatives)
at certain $\epsilon$, Eq.\ (\ref{expansion-ident}) holds only in
the limit $\omega_0 \rightarrow 0$. If the density of states
$g^{R(A)}_\epsilon$, $f^{R(A)}_\epsilon$, and the distribution
function were smooth functions, the quasi-static limit would hold
for any $\omega_0$; in this case all the quantities would simply
adiabatically depend on the AC potential $V_{ac}$. However, due to
a strong singularity at $\epsilon =\Delta$ of the density of
states and/or a sharp dependence of the distribution function for
low temperatures, the quasi-static picture breaks down for a
finite $\omega_0$, determined by the smallest scale of the
non-linearity. As a result both the tunnel current and the heat
flux for a finite frequency deviate strongly from the quasi-static
behavior.
In practice, the reservoirs are not perfect. In particular relaxation in the superconductor is still an open issue. 
This is a question that needs to be addressed separately. 
The present treatment gives the answers in the case of ideal reservoirs.

Consider the self-consistency equation for the order parameter.
Since $f_2 =0$ for $\varphi_S=0$, the self-consistency equation $
\Delta(\omega) = (\lambda/4) \int f^K_{\epsilon, \epsilon
-\hbar\omega}\, d\epsilon $ takes the form
\begin{equation}
\Delta (\omega) =\Delta _0 \sum_{k} J_k(2\alpha)2\pi \delta
(\omega -k\omega_0) \label{eq-Delta}
\end{equation}
which is the Fourier transform of Eq. (\ref{OP}) where
\[
\Delta_0({\bf r})=(\lambda/4) \int d\epsilon \left[
f^{R}_{\epsilon} -f^{A}_{\epsilon}\right]f_1(\epsilon)
\]
with $f_1(\epsilon)=\tanh (\epsilon/2T)$ is the order parameter
for zero AC field. This implies that the Eqs.
(\ref{chem-potential})--(\ref{t-dep-v}) are consistent. In
obtaining Eq. (\ref{eq-Delta}) we use
\begin{equation}
\sum_{n=-\infty}^{n=+\infty} J_{k+n}(t)J_{n}(z)=J_k(t-z)\ .
\label{J-ident3}
\end{equation}

\subsection{Charge and energy currents}\label{sec:theory-results}

For two tunnel-coupled electrodes, the charge current that flows
into the electrode $i$ is given by \cite{VHK05}
\begin{equation}
I(i) = -\frac{ie\pi \hbar\nu_i \Omega_i }{2}{\rm Tr}\,\left[\hat
\tau _3\hat I^K (i; t,t^\prime) \right]_{t =t^\prime} \ .
\label{charge-cons}
\end{equation}
whereas the heat current flowing into the electrode $i$ is
\begin{equation}
Q(i) =\frac{\pi \hbar^2\nu_i \Omega_i }{4} {\rm Tr}\!\left\{\!
\left[\frac{\partial}{\partial t}- \frac{\partial}{\partial
t^\prime} +\frac{2ie\varphi_i}{\hbar} \hat \tau _3
\!\right]\!\!\hat I^K (i; t,t^\prime)\! \right\}_{t =t^\prime} .
\label{energy-curr}
\end{equation}
Here $\nu_i$ is the normal-state density of states in the electrode
$i$, $\Omega_i$ and $\varphi_i $ are its volume and electric
potential. The collision integral $\hat I^K(i) $ in the electrode
that appears in Eqs.~(\ref{charge-cons}), (\ref{energy-curr})
contains contribution due to tunnelling from neighboring electrode
and the electron-phonon contribution, $I^K =I^K_t +I^K_{\rm e-ph}
$. The electron-electron interactions drop out from the energy
current because of the energy conservation. The energy flow into
the electrode can thus be separated into two parts. One part
containing $I^K_{\rm e-ph}$ is the energy exchange with the heat
bath (phonons). The other part contains the tunnel contribution
$I^K_t $ and is the energy current into the electrode through the
tunnel contact. The tunnel collision integral for the electrode 1
in contact with an electrode 2 has the form \cite{VHK05}
\begin{eqnarray}
\hat I^K_{t}(1)&=&i\eta_1 \left[ \hat g^R (2) \circ \hat
g^K(1)-\hat g^R (1) \circ \hat g^K (2)\right. \nonumber \\
&&\left. +\hat g^K (2) \circ \hat g^A (1)-\hat g^K(1) \circ \hat
g^A (2)\right]\ . \label{IKT-gen}
\end{eqnarray}
Here the arguments $i=1$ or 2 refer to the electrodes S or N. The
symbol $\circ$ is the convolution over the internal variables
\[
A(1)\circ B(2) =\int A(1;t_1, t^\prime)B(2;t^\prime , t_2)
dt^\prime\ .
\]
The factor
\[
\eta_i=[4\nu_i\Omega_i e^2 R_t]^{-1}
\]
parameterizes the tunnelling strength between the electrodes,
$R_t$ being the tunnel resistance. Since in the normal state
\begin{eqnarray*}
\hat g^{R(A)}_N(\epsilon;t_1,t_2)&=&\pm \hat \tau _3 \delta(t_1-t_2)\ ,
\\
\hat g^K_N(\epsilon;t_1,t_2)&=&2[f_{1}^N(\epsilon)\hat \tau _3 +
f_{2}^N(\epsilon)]\delta(t_1-t_2)\ ,
\end{eqnarray*}
the collision integral in the superconductor is
\begin{eqnarray}
\hat I^K_{t}(S)&=&i\eta_S \left\{ \hat \tau_3 \hat
g^K_S+\hat g^K_S \hat \tau_3 \right. \nonumber \\
&&\left. +2[\hat \tau_3 f_1^N+f_2^N]\hat g^A_S-2\hat
g^R_S[f_1^N\hat \tau_3+f_2^N]\right\} .\quad \label{collision}
\end{eqnarray}

The even and odd components of the distribution function
correspond to the absence of the AC potential. They are,
respectively,
\begin{eqnarray}
f_2^N (\epsilon) &=&-n_\epsilon +(1-n_{-\epsilon})\nonumber \\
&=&n_N(\epsilon+eU)-n_N(\epsilon -eU) \ ,\label{f2N}\\
f_1^N (\epsilon) &=&-n_\epsilon +n_{-\epsilon} \nonumber \\
&=&1- n_N( \epsilon+eU)-n_N(\epsilon -eU)\ ,\label{f1N}
\end{eqnarray}
for the normal lead, and
\begin{eqnarray}
f_2^S (\epsilon) &=&-n_\epsilon +(1-n_{-\epsilon})
=0 \ ,\label{f2S}\\
f_1^S (\epsilon) &=&-n_\epsilon +n_{-\epsilon} =1- 2n_S(
\epsilon)=\tanh\frac{\epsilon}{2T_S}\ , \label{f1S}
\end{eqnarray}
for the superconducting lead. Here $n_N(\epsilon)$ and
$n_S(\epsilon)$ are the Fermi functions with temperatures $T_N$
and $T_S$, respectively. The distributions in the superconductor
thus correspond to the zero-potential state.

As far as the NIS junction is concerned, consider first the charge
current into the superconductor defined by Eq.
(\ref{charge-cons}). The tunnel current in the frequency
representation becomes
\begin{widetext}
\begin{eqnarray}
I_S(\omega)&=&\frac{1}{4eR_t}\int d\epsilon\,
g_{\epsilon}\sum_{n,k}J_n(\alpha)J_{n-k}(\alpha) \left\{ 2\pi
\delta (\omega -k\omega_0) [f_1^S(\epsilon)+
f_2^S(\epsilon)-f_1^N(\epsilon+n\hbar\omega_0)-f_2^N(\epsilon+n\hbar\omega_0)]
\right.
\nonumber \\
&&\left. -2\pi \delta (\omega +k\omega_0) [f_1^S(\epsilon)-
f_2^S(\epsilon)-f_1^N(\epsilon -n\hbar\omega_0)+f_2^N(\epsilon
-n\hbar\omega_0)]\right\}\nonumber \\
&=&\frac{1}{2eR_t}\int d\epsilon\,
g_{\epsilon}\sum_{n,k}J_n(\alpha)J_{n-k}(\alpha)\nonumber \\
&& \times \left\{ 2\pi \delta (\omega -k\omega_0) [n_N(\epsilon
-eU +n\hbar \omega_0)-n_S(\epsilon)] +2\pi \delta (\omega
+k\omega_0) [ n_N(\epsilon -eU -n\hbar \omega_0)-n_S(\epsilon)]
\right\} \ . \label{I-total}
\end{eqnarray}
In Eq.~(\ref{I-total}) we use the relation $ \bar
g^{R(A)}_\epsilon =-g^{R(A)}_\epsilon $ for static functions, and
denote $ g_\epsilon \equiv (g^R_\epsilon -g^A_\epsilon)/2 $ the
ratio of the superconducting density of states to that in the
normal state, $g_{\epsilon}=N_S(\epsilon)/N_N $. The $\omega_0$
component of the current is
\begin{eqnarray}
I_{S\, \omega_0}(t)&=&\frac{\cos (\omega_0 t)}{eR_t} \int
d\epsilon\, g_{\epsilon}\sum_{n} J_n(\alpha)J_{n- 1}(\alpha)
\left[
f_1^S(\epsilon)-f_1^N(\epsilon+n\hbar\omega_0) \right]\nonumber \\
&=&\frac{\cos (\omega_0 t)}{eR_t} \int d\epsilon\,
g_{\epsilon}\sum_{n} J_n(\alpha)J_{n- 1}(\alpha)\left[
n_N(\epsilon +eU +n\hbar \omega_0) +n_N(\epsilon -eU + n\hbar
\omega_0) -2n_S(\epsilon)\right] \ . \label{current-ac}
\end{eqnarray}
\end{widetext}
The time averaged current takes the form
\begin{eqnarray}
\overline{I}_S&=&\frac{1}{2eR_t} \int d\epsilon\,
g_{\epsilon}\sum_{n}J_n^2(\alpha)\left[f_2^S(\epsilon)-
f_2^N(\epsilon+n\hbar\omega_0)\right]\nonumber \\
&=& \frac{1}{2eR_t} \int d\epsilon\,
g_{\epsilon}\sum_{n}J_n^2(\alpha)\nonumber \\
&&\times [n_N(\epsilon-eU+n\hbar \omega_0)- n_N(\epsilon+eU+n\hbar
\omega_0)] . \qquad \label{current-dc}
\end{eqnarray}
The terms with $f_2$ drop out of Eq.\ (\ref{current-ac}) due to
the property of the Bessel functions
\begin{equation}
J_{-n}=(-1)^n J_n \ . \label{J-ident1}
\end{equation}

Note that if we set $\omega_0=0$ the average current assumes the
{\em zero-ac-voltage} form
\[
I_S^{(0)}=\frac{1}{2eR_t} \int d\epsilon\, g_{\epsilon
}[n_N(\epsilon-eU)- n_N(\epsilon+eU)]\ .
\]
Indeed, when taking the limit $\omega_0 \rightarrow 0$ one should
keep in mind that the sum $ \sum _n J_n^2(\alpha) =1 $ [which is a
consequence of a more general relation Eq. (\ref{J-ident3})]
converges at $n\sim \alpha =eV_{ac}/\hbar \omega_0$. Therefore,
$\hbar n\omega _0 \sim eV_{ac} $ in Eq. (\ref{I-total}); thus one
has to put $eV_{ac}\rightarrow 0$ to neglect $n\hbar \omega_0$.
However, the true {\em static} expression is defined for
$\omega_0=0$ but $V_{ac}\neq 0$. According to Eq.
(\ref{expansion-ident}), it has the energy-shifted density of
states $g_{\epsilon \pm eV_{ac}}$ and the distribution functions
$f_1^S(\epsilon \pm eV_{ac})$, $f_2^S(\epsilon \pm eV_{ac})$. This
{\em static} limit (i.e., $\omega_0=0$ and $V_{ac}\neq 0$) is
indeed obtained from Eq.~(\ref{I-total}) using Eq.\
(\ref{expansion-ident}). Making shifts of the integration variable
we find
\begin{equation}
I_S^{\rm static}=\frac{1}{2eR_t} \int d\epsilon\, g_{\epsilon}[
f_2^S(\epsilon)-f_2^{N\ {\rm static}}(\epsilon)]
\end{equation}
where $f_2^S=0$ and
\[
f_2^N=n_N(\epsilon +eU-eV_{ac})-n_N(\epsilon -eU+eV_{ac})
\]
which corresponds to the total voltage $U-V_{ac}$, according to
Eq. (\ref{f2N}).

The heat current that flows into the superconducting lead can be
calculated with help of Eqs. (\ref{t-dep-u}) and (\ref{t-dep-v}).
We find in the frequency representation
\begin{eqnarray}
&&\!\!-\hbar\left[\left(\frac{\partial}{\partial t}-
\frac{\partial}{\partial t^\prime} +\frac{2ie\varphi_S}{\hbar}
\right)g^{R(A)}(t,t^\prime)\right]_{\epsilon
,\epsilon-\hbar\omega}
 \label{kernel1}\\
&=&\!\!4\pi i\sum_{n,k}\delta (\omega
-k\omega_0)J_n(\alpha)J_{n-k}(\alpha)(\epsilon -n\hbar\omega
_0)g^{R(A)}_{\epsilon -n\hbar\omega_0 } ,\nonumber \\
 &&\!\! -\hbar \left[\left( \frac{\partial}{\partial
t}- \frac{\partial}{\partial t^\prime} +\frac{2ie\varphi_S}{\hbar}
\right)g^K(t,t^\prime)\right]_{\epsilon ,\epsilon-\hbar\omega} \label{kernel2} \\
&=&\!\! 4\pi i\sum_{n,k} \delta (\omega
-k\omega_0)J_n(\alpha)J_{n-k}(\alpha)(\epsilon -n\hbar\omega
_0)\nonumber \\
&&\!\! \times [g^{R}_{\epsilon -n\hbar\omega_0 }-g^{A}_{\epsilon
-n\hbar\omega_0 }] \left[f_{1}(\epsilon -n\hbar\omega _0) +
f_{2}(\epsilon -n\hbar\omega _0)\right] ,\nonumber
\end{eqnarray}
and similarly for $\bar g$ with the substitutions $g\rightarrow
\bar g$, $\varphi_S \rightarrow - \varphi_S$, $\omega_0
\rightarrow - \omega_0$, and $f_2 \rightarrow -f_2$. Here the
distribution functions again correspond to zero AC potential.

Shifting the energy variable under the integral, the average heat
current into the superconductor becomes
\begin{eqnarray}
\overline{Q}_S &=& \frac{1}{2e^2 R_t} \int
 \epsilon g_{\epsilon}\sum _nJ_n^2(\alpha)\nonumber \\
&& \times \left[f_1^S(\epsilon )-f_1^N(\epsilon+n\hbar
\omega_0)\right]\, d\epsilon \ . \quad \label{heatcurrent2}
\end{eqnarray}
The heat current Eq. (\ref{heatcurrent2}) is even in $\omega_0$.
For $\omega_0 =0$ Eq.~(\ref{heatcurrent2}) formally goes over into
\[
Q_S^{(0)}=\frac{1}{2e^2R_t}\int \epsilon g_\epsilon \left[
f_1^S(\epsilon)-f_1^N(\epsilon)\right]\, d\epsilon
\]
with $f_1^N$ and $f_1^S$ from Eqs.\ (\ref{f1N}), (\ref{f1S}). This
is the {\em zero-ac-voltage} result.

The {\em static} expression is obtained from Eqs.
(\ref{expansion-ident}), (\ref{kernel1}), and (\ref{kernel2})
\begin{equation}
Q_S^{\rm static}=\frac{1}{2e^2R_t}\int \epsilon g_\epsilon \left[
f_1^S(\epsilon)-f_1^{N\ {\rm static}}(\epsilon)\right]\, d\epsilon
\label{Qstat}
\end{equation}
where $f_1^{N{\rm static}}(\epsilon)$ corresponds to the total
voltage $U-V_{ac}$,
\[
f_1^{N\ {\rm static}}(\epsilon)= 1-n_N(\epsilon
+eU-eV_{ac})-n_N(\epsilon -eU+eV_{ac})\ .
\]
This should be compared to Eq. (\ref{f1N}).

It is also interesting to define the {\em quasi-static} regime,
which is obtained by averaging the static heat flux
$Q_{\text{S}}^{\text{static}}$ over the sinusoidal voltage cycle
with $V_{\text{ac}}\rightarrow V_{\text{ac}}\cos{\omega_0t}$. It
does not coincide with the static expression due to the voltage
oscillations. This quasi-static regime corresponds to the
classical limit occurring at small frequencies, for which the
photon energy $\hbar \omega_0$ is much smaller than the energy
scale over which the non-linearity of the I-V curve occurs
\cite{tucker,tucker2}. In the system under investigation such
energy scale is set by the temperature or by the width of the
superconducting DOS peak near the gap energy, which smear the
sudden current onset occurring at the superconductor gap. As it
will be confirmed in Sec.~\ref{sec:result}, the quasi-static
regime occurs for $\hbar \omega_0\ll k_{\text{B}}T$.

We now consider the heat current flowing out of the normal electrode:
\begin{equation}
Q^{\rm out}_N =Q_S-(\varphi _N -\varphi_S)I_S \label{Q-out} ,
\end{equation}
where $I_S$ is the tunnel charge current reported in Eq. (\ref{I-total}).
Note that the heat extracted from the normal
electrode and the heat entering the superconducting lead differ by
the energy absorbed at the NIS interface where the potential drops
by $\varphi_N -\varphi_S$.
The time-average heat current is
\begin{equation}
\overline{Q}^{\rm out}_N = \overline{Q}_S- U\overline{I}_S -P_{ac}
\label{heat-current-total}
\end{equation}
where $P_{ac}=\overline{ V_{\text{ac}}\cos(\omega_0 t)I_{S\, \omega_0}(t)}$ is
the average AC power absorbed at the NIS contact,
\begin{eqnarray}
P_{ac} &=&- \frac{V_{\text{ac}}}{2eR_t} \int g_{\epsilon
}\sum_{n} J_n(\alpha)J_{n- 1}(\alpha) \nonumber \\
&& \times \left[f_1^S(\epsilon)- f_1^N(\epsilon +
n\omega_0)\right]\, d\epsilon
 \ . \; \label{Pac}
\end{eqnarray}
Note that $P_{ac}$ is finite both in the static case
($\omega_0\equiv 0$) and in the quasi-static regime (small but
finite $\omega_0$).

\section{Results and discussion}
\label{sec:result} We shall now discuss how the heat current
depends on the various parameters of the system. This can be done
by numerically evaluating the expressions given in the previous
section. In the following we shall assume parameters typical of
aluminum (Al) as S material,
with critical temperature $T_c$=1.19~K. We assume the
superconducting gap to follow the BCS relation
$\Delta_0=1.764k_BT_c$ and choose
\[
N_S(\epsilon)=|\text{Re}[(\epsilon+i\Gamma)/
\sqrt{(\epsilon+i\Gamma)^2-\Delta^2}]|\ ,
\]
where $\Gamma$ is a smearing parameter which accounts for
quasiparticle states within the gap \cite{VHK05,Dynes,pekola2004}.
We will use $\Gamma=10^{-4}\Delta_0$, as experimentally verified
in Ref.~\onlinecite{pekola2004}. Finally, we shall always assume
the N and S electrodes to be at the same temperature, i.e.,
$T_S=T_N\equiv T$.

\begin{figure}[t!]
\includegraphics[width=\columnwidth,clip]{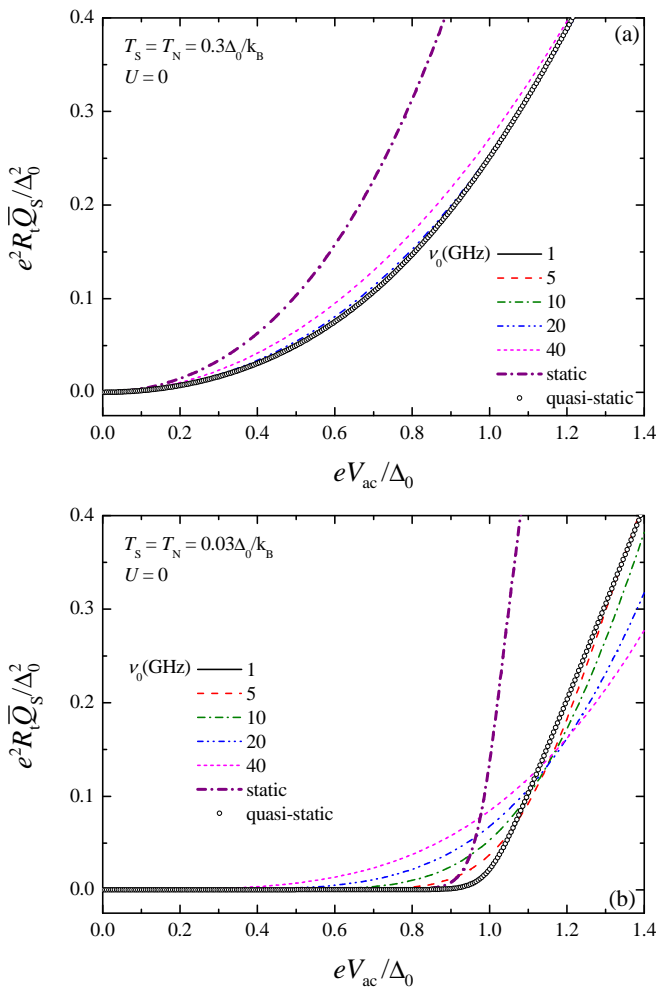}
\caption{(color online) Normalized time-average heat current into the S
electrode $\overline{Q}_{\text{S}}$ as a
function of the amplitude of the AC voltage $V_{\text{ac}}$ for
different values of $\nu_0$ at (a) high ($T=0.3\Delta_0/k_{\text{B}}$)
and (b) low ($T=0.03\Delta_0/k_{\text{B}}$) temperature.
The static case and the quasi-static limit are plotted for comparison.
Note that in (a) the curves relative to $\nu_0=$10, 5 and 1 GHz coincide
with the quasi-static one.}
\label{fig4}
\end{figure}

For the sake of definiteness, let us first consider the situation
in which no bias is applied to the normal island ($U=0$). In
Figs.~\ref{fig4}(a) and (b) the time-averaged heat current
entering the S electrode ($\overline{Q}_{\text{S}}$),
Eq.~(\ref{heatcurrent2}), is plotted as a function of the AC
voltage at, respectively, large ($T=0.3\Delta_0/k_{\text{B}}$) and
small ($T=0.03\Delta_0/k_{\text{B}}$) temperatures. The various
curves refer to different values of frequency
$\nu_0=\omega_0/2\pi$, and calculations were performed up to $\nu_0=40$ GHz,
corresponding roughly to the value of the superconducting gap
($\Delta_0/h\simeq43.7$ GHz). We note that a driving frequency
corresponding to $2\Delta_0$ would lead to breaking up of the
Cooper pairs.
For a comparison we have included static
(i.e., $\nu_0= 0$), Eq. (\ref{Qstat}), and quasi-static regimes.
For large temperatures [i.e., $T=0.3\Delta_0/k_{\text{B}}$,
Fig.~\ref{fig4}(a)], the heat current $\overline{Q}_{\text{S}}$ is
a monotonic, nearly parabolic, function of $V_{\text{ac}}$ for all
values of frequency. The first observation is that the static heat
current is always larger than the heat current at
finite frequency. On the one hand, it is obvious that the
quasi-static curve is below the static one, the former being just
an average over a cycle of the static limit (see
Sec.~\ref{sec:theory-results}).
\begin{figure}[t!]
\includegraphics[width=\columnwidth,clip]{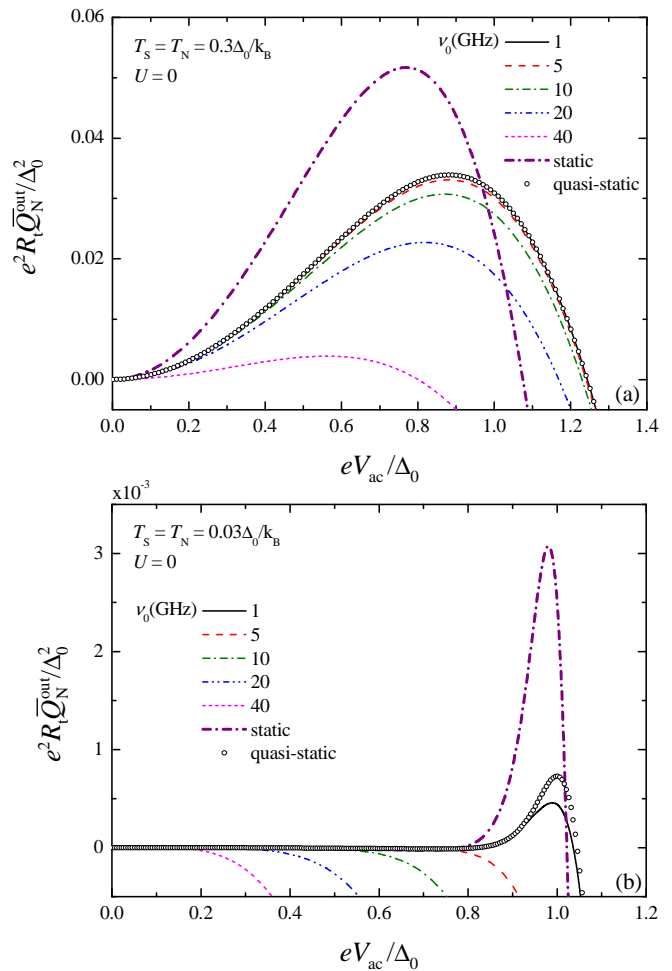}
\caption{(color online) Normalized time-averaged heat current out of the
N metal $\overline{Q}_{\text{N}}^{\text{out}}$ as a
function of the amplitude of the AC voltage $V_{\text{ac}}$ for different
values of frequency $\nu_0$ at (a) high ($T=0.3\Delta_0/k_{\text{B}}$) and (b)
low ($T=0.03\Delta_0/k_{\text{B}}$) temperature. The static case and
the quasi-static limit are plotted for comparison. At high temperature the
quasi-static
limit is a good approximation at frequencies as low as 1 GHz.}
\label{fig2}
\end{figure}
On the other hand, the
photon-assisted heat current is always larger than quasi-static
characteristic. To be more precise, the heat current monotonically
decreases by decreasing frequency, eventually reaching the
quasi-static limit for small enough $\nu_0$ (note that the curves
relative to $\nu_0=1,5,10$ GHz are indistinguishable from the
quasi-static one). This means that photon-assisted processes give
rise to an enhancement of the heat current entering S with respect
to the quasi-static situation, though remaining well below static
values. Such enhancement reflects the increase in current due to
photon-assisted processes~\cite{tien}: electrons are excited to
higher energy states, thus favoring tunneling above the gap. Of
course, such mechanism is more effective for small temperatures.
In such a case [i.e., $T=0.03\Delta_0/k_{\text{B}}$,
Fig.~\ref{fig4}(b)], indeed, static and quasi-static curves
present an activation-like behavior, with a switching voltage of
$V_{\text{ac}}\simeq 0.9\Delta_0/e$ and $V_{\text{ac}}\simeq
1.0\Delta_0/e$, respectively, and thereby increasing almost
linearly. Photon-assisted $\overline{Q}_{\text{S}}$ increases more
smoothly as compared with the quasi-static case, which is
approached by decreasing $\nu_0$.

We now consider the heat current extracted from the N electrode
$\overline{Q}_{\text{N}}^{\text{out}}$, which differs from
$\overline{Q}_{\text{S}}$ by the AC power $P_{\text{ac}}$ (for
$U=0$) absorbed by the NIS contact [see
Eqs.~(\ref{heat-current-total}), (\ref{Pac})]. In
Figs.~\ref{fig2}(a) and (b) we plot
$\overline{Q}_{\text{N}}^{\text{out}}$ as a function of
$V_{\text{ac}}$ for several frequencies for large and small
temperature, respectively. The effect of $P_{\text{ac}}$ on the
behavior of the heat current is very strong, giving rise to a
maximum located around $V_{\text{ac}}=\Delta_0/e$, and to a sign
change. By increasing $V_{\text{ac}}$ the heat flow out of N
increases up to the maximum and thereafter rapidly decreases to
negative values (heat current enters the N electrode). For reasons
given above, the maximum quasi-static heat current is always
smaller than the maximum of the static one. Another effect of
$P_{\text{ac}}$ is that, in this case, the photon-assisted heat
current is smaller than the quasi-static characteristic. In particular,
the heat current monotonically increases by
decreasing frequency, eventually reaching the quasi-static limit
for small enough $\nu_0$. Moreover, by increasing the frequency
the maximum of $\overline{Q}_{\text{N}}^{\text{out}}$ moves toward
smaller values of $V_{\text{ac}}$. While at large temperatures
$\overline{Q}_{\text{N}}^{\text{out}}$ remains positive (implying
heat extraction from the N electrode) also for frequencies
slightly above 40 GHz [see Fig.~\ref{fig2} (a)], at low
temperatures the minimum frequency for positive
$\overline{Q}_{\text{N}}^{\text{out}}$ is drastically reduced (by
about one order of magnitude) [see Fig.~\ref{fig2} (b)]. This
clearly proves that photon-assisted tunneling is detrimental as
far as heat extraction from the N electrode is concerned.
\begin{figure}[t!]
\includegraphics[width=\columnwidth,clip]{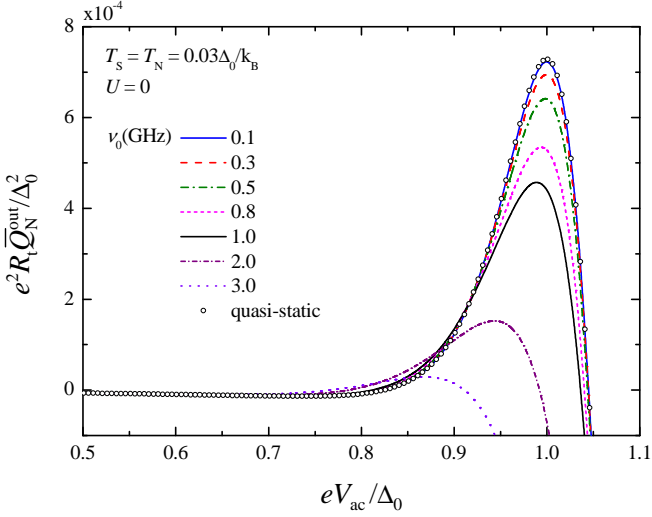}
\caption{(color online) The same as for Fig.~\ref{fig2} (b) for
frequencies in a smaller range. At low temperature the
quasi-static limit is a good approximation for frequencies as low
as 0.1 GHz.} \label{fig3}
\end{figure}
Analogously to what happens for the charge current~\cite{tien},
the approach to the quasi-static limit depends on temperature.
Indeed, as already mentioned in Sec.~\ref{sec:theory-results}, the
quasi-static regime occurs at $\hbar\omega_0 \ll k_{\text{B}}T$.
The curve relative to 1 GHz differs, with respect to the
quasi-static one at its maximum, by less than 0.1\% at
$T=0.3\Delta_0/k_{\text{B}}$, and by about 50\% at
$T=0.03\Delta_0/k_{\text{B}}$, where
$k_{\text{B}}T\sim\hbar\omega_0$. Figure~\ref{fig3} shows the
time-average $\overline{Q}_{\text{N}}^{\text{out}}$ versus
$V_{\text{ac}}$ at low temperature calculated for frequencies in a
smaller range. As it can be clearly seen, the quasi-static curve
appears to be a good approximation for $\nu_0=0.1$ GHz.

\begin{figure}[t!]
\includegraphics[width=\columnwidth,clip]{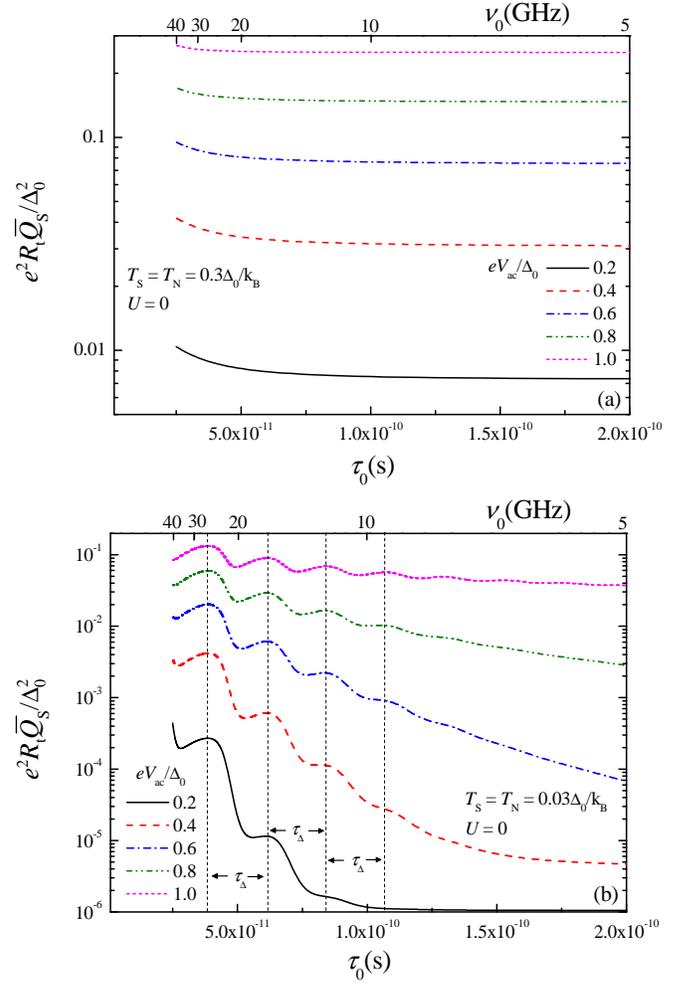}
\caption{(color online) Normalized time-average heat current into
the S electrode $\overline{Q}_{\text{S}}$ as a function of the
period of the oscillations $\tau_0=1/\nu_0$ for various values of
$V_{\text{ac}}$ at (a) $T=0.3\Delta_0/k_{\text{B}}$ and (b)
$T=0.03\Delta_0/k_{\text{B}}$. The distance between the relative
maxima at low temperature [panel (b)] is
$\tau_{\Delta}=h/\Delta_0$. }
\label{fig6}
\end{figure}
\begin{figure}[t!]
\includegraphics[width=\columnwidth,clip]{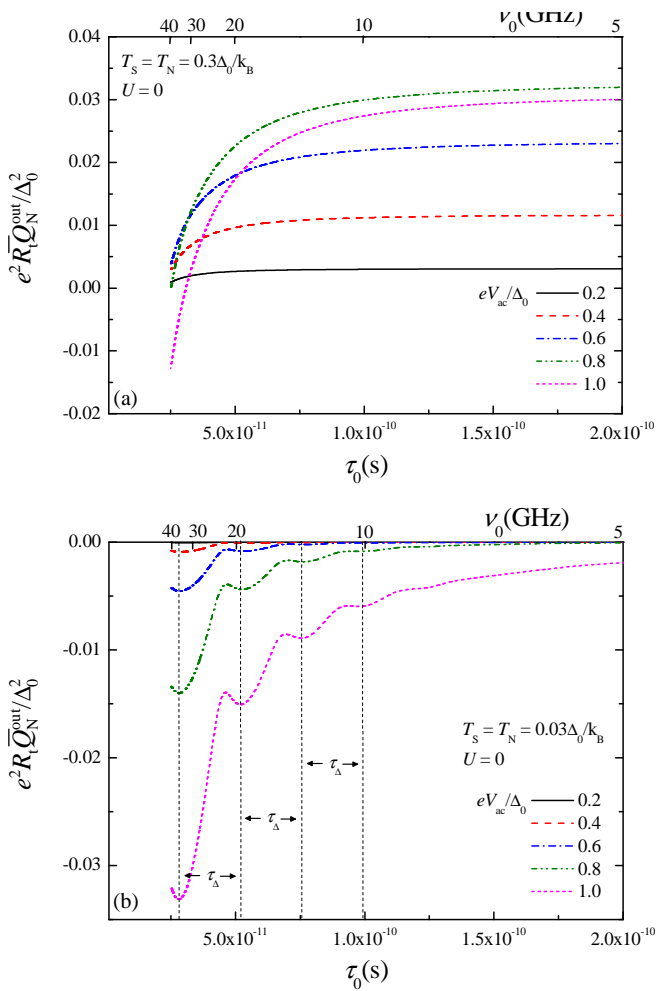}
\caption{(color online) Normalized time-average heat current out
of the N electrode $Q_{\text{N}}^{\text{out}}$ as a function of
the period of the oscillations $\tau_0=1/\nu_0$ for several values
of $V_{\text{ac}}$ at (a) $T=0.3\Delta_0/k_{\text{B}}$ and (b)
$T=0.03\Delta_0/k_{\text{B}}$. At low temperature [panel (b)]
the distance between relative minimums turn out to be
$\tau_{\Delta}=h/\Delta_0$. In (b) the curve relative to
$V_{\text{ac}}=0.2\Delta_0/e$ is consistent, in the scale of the
figure, with zero. } \label{fig5}
\end{figure}

It is now interesting to analyze the behavior of dynamic heat
transport in the NIS junction for fixed amplitude of the AC
voltage by plotting the heat currents as a function of the period
of oscillations $\tau_0=1/\nu_0$. This is shown in
Figs.~\ref{fig6} and \ref{fig5} for $\overline{Q}_{\text{S}}$ and
$\overline{Q}_{\text{N}}^{\text{out}}$, respectively. Here we set
$U=0$.  Both for large [$T=0.3\Delta_0/k_{\text{B}}$, see
Fig.~\ref{fig6}(a)] and small [$T=0.03\Delta_0/k_{\text{B}}$, see
Fig.~\ref{fig6}(b)] temperatures the heat current
$\overline{Q}_{\text{S}}$ presents an overall decrease with
$\tau_0$, for all values of $V_{\text{ac}}$. At small
temperatures, however, the heat current shows an additional
structure consisting of superimposed oscillations due to
photon-assisted processes, which tend to disappear for large
values of $\tau_0$ (small frequencies), i.e., approaching the
quasi-static limit. Notably, the relative maxima turn out to be
equally spaced by the time scale related to the superconducting
gap, $\tau_{\Delta}=h/\Delta_0$. In addition we found that, at even
lower temperatures, also the relative maxima are equally spaced
by $\tau_{\Delta}$.

Though presenting an overall enhancement with $\tau_0$, the
behavior of the time-average heat current extracted from the N
electrode $\overline{Q}_{\text{N}}^{\text{out}}$ is qualitatively
similar to that of $\overline{Q}_{\text{S}}$ (see Fig.~\ref{fig5};
note that the vertical axis is linear in this case). Note that for
large temperatures $\overline{Q}_{\text{N}}^{\text{out}}$ remains
positive for most of the frequency range considered, even for
$V_{\text{ac}}=\Delta_0/e$ [see Fig.~\ref{fig5}(a)]. For small
temperatures [see Fig.~\ref{fig5}(b)], however, the heat current
is negative over nearly the whole time range. The additional
structure, in this case, shows equal spacing (of magnitude
$\tau_{\Delta}$) between the relative minimums, since these
correspond to maximum heat absorption by S [maxima in
Fig.~\ref{fig6}(b)].

\begin{figure}[t!]
\includegraphics[width=\columnwidth,clip]{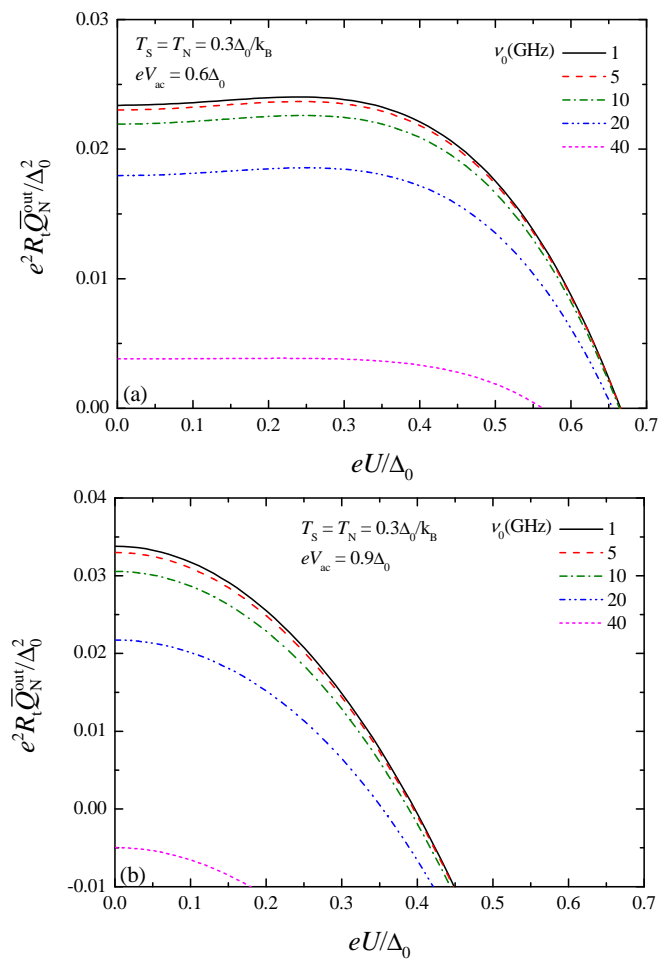}
\caption{(color online) Normalized time-average heat current
exiting the N electrode $\overline{Q}_{\text{N}}^{\text{out}}$ as
a function of the DC voltage $U$ for various values of $\nu_0$ at
(a) $V_{\text{ac}}=0.6\Delta_0/e$ and (b)
$V_{\text{ac}}=0.9\Delta_0/e$. In (a) and (b) the calculations
were performed by setting the temperature at
$T=0.3\Delta_0/k_{\text{B}}$.} \label{fig7}
\end{figure}
We now turn to the effect of a finite DC voltage $U$ combined with
an AC modulation on the heat current exiting the N electrode. In
Figs.~\ref{fig7}(a) and (b), the time-average
$\overline{Q}_{\text{N}}^{\text{out}}$ at large temperatures
($T=0.3\Delta_0/k_{\text{B}}$) is plotted as a function of $U$ for
several values of frequency at $V_{\text{ac}}=0.6\Delta_0/e$ and
$V_{\text{ac}}=0.9\Delta_0/e$, respectively. Figure~\ref{fig7}(a)
shows that $\overline{Q}_{\text{N}}^{\text{out}}$ is nearly
constant having a weak maximum around $U\simeq 0.3\Delta_0/e$
almost independently of the frequency, and rapidly decreasing
thereafter. By inspecting Fig.~\ref{fig2}(a) it clearly appears
that, at $T=0.3\Delta_0/k_{\text{B}}$,
$\overline{Q}_{\text{N}}^{\text{out}}$ is maximized around
$V_{\text{ac}}\approx 0.9\Delta_0/e$, so it seems that a finite
value of $U$ just adds to the AC voltage making the heat current
to move along the voltage characteristic similarly to the pure AC
case. Furthermore, we note that the addition of a static DC
potential to an AC modulation is not able to recover the maximum
value the heat current can achieve with only the AC voltage
biasing. A confirmation of this is given in the plots displayed in
Fig.~\ref{fig7}(b) which are relative to a value of
$V_{\text{ac}}=0.9\Delta_0/e$. For such an AC voltage biasing
$\overline{Q}_{\text{N}}^{\text{out}}$ does not present a constant
part, and the addition of $U$ turns out to only suppress the
time-average heat current. Moreover, an increase of frequency
$\nu_0$ causes a reduction of
$\overline{Q}_{\text{N}}^{\text{out}}$, even to negative values.

\begin{figure}[t!]
\includegraphics[width=\columnwidth,clip]{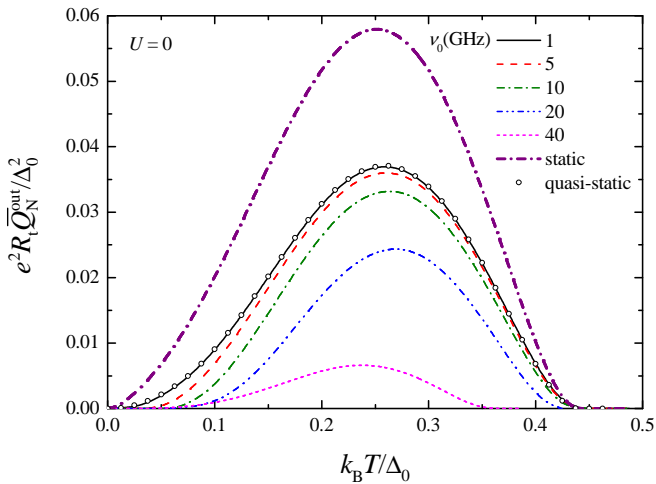}
\caption{(color online) Normalized time-average heat current
out of the N electrode $\overline{Q}_{\text{N}}^{\text{out}}$,
maximized with respect to $V_{\text{ac}}$, as a function of
temperature $T$ at $U=0$ for different values of frequency $\nu_0$.
The static case and the quasi-static limit are plotted for
comparison.} \label{fig8}
\end{figure}

We finally plot in Fig.~\ref{fig8} the maximum value of
$\overline{Q}_{\text{N}}^{\text{out}}$, obtained by spanning over
$V_{\text{ac}}$, as a function of $T$ for several values of
frequency. For every $\nu_0$ the time-average
$\overline{Q}_{\text{N}}^{\text{out}}$ is a bell-shaped function
presenting a maximum around $T\simeq0.25\Delta_0/k_{\text{B}}$
(similarly to what happens in the static \cite{giazotto06} as well
as in the quasi-static limit), which is gradually suppressed upon
enhancing the frequency. By increasing the frequency the curves
slightly shrink, thus reducing the temperature interval of
positive heat current. Moreover, the position of the maxima
tends to move to higher temperatures for intermediate frequencies
(i.e., for $\nu_0$ below $\simeq 20$ GHZ), while they tend to move
to lower temperatures in the higher range of frequencies (see
for example the curve corresponding to $\nu_0=40$ GHz in Fig.
\ref{fig8}).

\section{Conclusions}
\label{sec:conc} In this paper we have calculated the heat
currents in a normal/superconductor tunnel junction driven by an
oscillating bias voltage in the photon-assisted tunneling regime.
We have found that the maximum heat extracted from the normal
electrode  decreases with increasing driving frequency. We checked
that for small frequencies ($\hbar\omega_0\ll k_{\text{B}}T$) the
photon-assisted heat current approaches the quasi-static limit,
the latter being obtained by averaging the static heat current
over a sinusoidal voltage cycle (relevant for sub-GHz
frequencies). The suppression of the heat current by
photon-assisted processes can be imputed to the AC power,
dissipated at the tunnel contact, which is enhanced in the quantum
regime with respect to the quasi-static limit. On the contrary,
the heat current entering the superconducting electrode slightly
increases with increasing frequency. We also found that, for
small temperatures, the heat current as a function of the inverse
of frequency presents an additional structure consisting of
superimposed oscillations with a period corresponding to the time
scale derived from the superconducting gap,
$\tau_{\Delta}=h/\Delta_0$.

We want finally to briefly comment onto some implications of the above results 
for practically realizeable systems. We refer, for instance, to AC-driven NIS electron refrigerators operating in the regime of Coulomb blockade 
which were theoretically investigated in Ref. \cite{pekola07a}, and experimentally demonstrated in Ref. \cite{saira}. More in particular, it was shown in Ref. \cite{pekola07a} that both the heat current flowing out the N island and the minimum achievable electron temperature depend on the frequency of the gate voltage 
as well as on the bath temperature. Our results may thus suggest the proper operating conditions in terms of frequencies and bath temperatures 
in order for photon-assisted tunneling not to suppress the heat current in these systems.
In other words, they allow to predict a suitable range of parameters which keep the system in the quasi-static limit.

\section{Acknowledgments}
We thank R. Fazio for careful reading of the manuscript. Partial
financial support by the Russian Foundation for Basic Research grant
06-02-16002, from Academy of Finland, and from the EU funded
NanoSciERA ``NanoFridge'' and RTNNANO projects is
acknowledged.

\appendix
\section{Static limit}
\label{app}

To prove Eq. (\ref{expansion-ident}) we use that $\Phi (x)$ is
analytic, therefore
\begin{eqnarray*}
\Phi(\epsilon -n\hbar \omega_0)e^{-in\omega_0t}=\sum _{k=0}^\infty
\frac{d^k \Phi}{d\epsilon
^k}\frac{(-n\hbar\omega)^k}{k!}e^{-in\omega_0t}\\
=\Phi\left(\epsilon -i\hbar\frac{\partial }{\partial
t}\right)e^{-in\omega_0t}
\end{eqnarray*}
We next perform the inverse Fourier transform of the right-hand
side of Eq. (\ref{expansion-ident})
\begin{eqnarray*} \Phi
(\epsilon +\mu_s (t))=\sum _{n,k}J_n(\alpha)J_{n-k}(\alpha)
\Phi(\epsilon-n\hbar\omega_0 )e^{-ik\omega _0 t}\\
=\sum_n J_n(\alpha)\Phi(\epsilon -n\hbar\omega_0)e^{-in\omega_0t}
e^{\frac{i}{\hbar}\int_0^t eV_{ac}\cos \omega_0 t^\prime\, dt^\prime}\\
=e^{\frac{i}{\hbar}\int_0^t eV_{ac}\cos \omega_0 t^\prime\,
dt^\prime}\sum_n \Phi\left(\epsilon -i\hbar\frac{\partial
}{\partial
t}\right)J_n(\alpha)e^{-in\omega_0t}\\
=e^{\frac{i}{\hbar}\int_0^t eV_{ac}\cos \omega_0 t^\prime\,
dt^\prime}\Phi\left(\epsilon -i\hbar\frac{\partial }{\partial
t}\right)e^{-\frac{i}{\hbar}\int_0^t eV_{ac}\cos \omega_0
t^\prime\, dt^\prime}
\end{eqnarray*}
which indeed is $\Phi (\epsilon -eV_{ac}\cos \omega_0 t)$. Here we
use the expansion Eq. (\ref{expansion1}) twice.

\end{document}